\documentclass[12pt]{iopart}
\usepackage{iopams}  
\usepackage{xcolor}
\usepackage[round]{natbib}
\usepackage{datetime}
\newdateformat{monthyeardate}{\monthname[\THEMONTH], \THEYEAR}
\usepackage{graphicx}
\usepackage[textheight = 8.5in]{geometry}
\usepackage{multirow} 
\usepackage[singlelinecheck=false]{caption}
\usepackage [english]{babel}
\usepackage [autostyle, english = american]{csquotes}
\MakeOuterQuote{"}

\usepackage{scalerel}
\usepackage{array}
\newcolumntype{C}[1]{>{\centering\arraybackslash}m{#1}}

\usepackage{caption}
\usepackage{subcaption}
\usepackage{tikz}
\usepackage{tkz-tab}
\usetikzlibrary{knots,calc}  
\usepackage{pgfplots}
\pgfplotsset{compat=1.17}

\begin{document}

\title[Robustness of DNN models]{Experimental and Numerical Investigations on Robustness of a Deep Neural Network-based Multi-class Classification Model of CT Images with respect to Image Noise}

\author{Yuting Peng$^1$, Chenyang Shen$^2$, Yesenia Gonzalez$^2$, Yin Gao$^2$, Xun Jia$^1$}

\address{1. Department of Radiation Oncology and Molecular Radiation Sciences, Johns Hopkins University, Baltimore, MD, USA}
\address{2. Department of
Radiation Oncology, University of Texas Southwestern Medical Center, Dallas, TX, USA}
\vspace{10pt}
\begin{indented}
\item[]\monthyeardate\today
\end{indented}

\begin{abstract}

\textbf{Objective:} Robustness of Deep Neural Networks (DNNs) is an important aspect to consider for their clinical applications. This work examined robustness issue for a DNN-based multi-class classification model via comprehensive experimental and simulation studies. 

\textbf{Approach:} We constructed a DNN-based multi-class classification model that classifies an axial CT image as one of the four body sites of brain \& neck (BN), chest (C), abdomen \& pelvis (AP), and leg \& foot (LF). The model was trained with whole-body CT images of 37 patients each scanned once and 10 scans of a whole-body phantom with different mAs levels to achieve a F1 score of 99.7\% averaged over four classes on a testing dataset. We evaluated robustness of the model against noise perturbations under different mAs levels using simulated CT noisy images based on noise power spectrum (NPS) and experimental CT images acquired in repeated scans of the orange-man phantom. To quantify robustness of the trained model, we defined Successful Attack Rate (SAR) as the ratio of predictions changed with noises and Confusion Matrix for Robustness (CMR) that represents the percentage of a predicted class without noise being predicted to different classes under noise. Besides, we repeatedly trained the model for 100 times using exactly the same training procedure, hyper-parameter settings, as well as the training and validation dataset to investigate robustness of models due to inherent randomness in the training process. Finally, to improve model robustness, we employed an adaptive training scheme and demonstrated its effectiveness.

\textbf{Main results:} Results in experimental study agreed with those of simulation study in terms of quantitatively evaluating model robustness. As mAs levels decreased, the DNN-based model showed worse robustness with respect to amplified CT noises. At 25 mAs level, SAR was 18\%, 57.3\%, 0\%, and 28.7\%, for the BN, C, AP, and LF sites, respectively. Among 100 repeatedly trained DNN models with exactly the same setting, 14 showed robustness issue to various degrees at 25 mAs. The adaptive training scheme was effective in terms of improving model robustness, such that within two iterations model robustness issue was mitigated for both mAs levels.

\textbf{Significance:} Experimental and numerical studies demonstrated robustness issue of a DNN-based multi-class classification model. The discoveries highlight the needs for evaluating and ensuring DNN model robustness.

\end{abstract}

%
%
\vspace{2pc}
\submitto{\PMB}
%
\maketitle
%

\section{Introduction}

Deep learning (DL) methods have attracted increasingly interests in medical domains in recent years. Employing a large-scale hierarchical multi-layer structure often in the form deep neural networks (DNNs), DL methods are extremely capable of learning feature embeddings of the data to allow state-of-the-art performance in representing the distribution followed by the data for various purposes \citep{kim2019deep,shen2020introduction}. With its success, DNN models have been applied to a spectrum of tasks including, but not limited to, medical image reconstruction and processing \citep{jiang2019augmentation,han2017mr}, disease diagnosis \citep{esteva2017dermatologist,lakhani2017deep}, outcome prediction of therapeutic approaches \citep{ewbank2020quantifying, men2019deep}, radiotherapy treatment planning \citep{nguyen20193d,kontaxis2020deepdose,shen2020operating,shen2019intelligent,gao2022modeling}, and quality assurance \citep{nyflot2019deep,tomori2018deep}.

Despite these successes, understandings of DNNs' mathematical properties are few and far. Yet, it is desired to understand the DNN's properties from theoretical perspective for practical applications, as they elucidate strengths and challenges of DNN models and help ensuring validity of applications. One particular aspect of the DNN is its stability and robustness, which refers to the sensitivity of a DNN model's output with respect to small perturbations to the model's input. A DNN consists of a large number of neuron units connected under a certain structure. While the mathematical operation of each unit is simple, the end-to-end mapping established from the input to the output is extremely complex. Understanding stability of this complex mapping is critical, because in real-world applications, small perturbations to the input data are inevitable, e.g. in the form of noise, and a useful  model have to be able to tolerate the noise perturbations. 

Along with developments of DNN models, there have been studies demonstrating the stability and robustness concern in different applications \citep{DBLP:journals/corr/abs-2112-00639,laugros2019adversarial,eykholt2018robust,madry2017towards,akhtar2018threat,yuan2019adversarial}. Under naturally-induced image corruptions or alternations, models sometimes may not preserve performance to confidently output correct results. Specific to the medicine domain, \citet{finlayson2018adversarial} demonstrated the existence of perturbations to input data that can affect outputs of three representative medical DL classifiers for the classification of diabetic retinopathy from retinal fundoscopy, pneumothorax from chest-xray, and melanoma from dermoscopic photographs. \citet{shen2020robustness} observed the similar behavior for a DL-based classifier for lung-nodule classification from CT images \citep{shen2020robustness,gao2021improving}. Stability issue has also been observed for medical image processing applications. \citet{antun2020instabilities} investigated possible reasons for network instabilities in image reconstruction tasks. \citet{wu2020stabilizing,wu2021drone} developed a method integrating Analytic Compressed Iterative Deep based reconstruction method and Dual-domain Residual-based Optimization Network to eliminate the instabilities. Researchers have also found that enforcing sparsity in network parameters is an effective approach to improve model robustness \citep{hoefler2021sparsity,ozturk2021convolutional}.

Previous investigations on the robustness of DNN models have been performed mostly via numerical approaches. In \citep{shen2020robustness,gao2021improving}, CT noise signals were sampled based on noise amplitude and a realistic noise power spectrum. \citet{cui2021deattack} proposed an evolution based attack method to evaluate the robustness of medical image segmentation. A recent study reported a framework for testing robustness of machine learning-based classifiers\citep{jpm12081314}. While these efforts demonstrated the model stability issue, it is desirable to have experimental studies to confirm the findings. Nonetheless, it is a practically challenging problem to assess model robustness in experimental settings. A task like this would require repeated data acquisition of the same model input under different noise realizations. Not only is this a tedious task, in some cases, this is not even feasible for ethic considerations. For instance, it is impossible to rescan a patient with CT to evaluate stability of a DNN model for CT-based disease classification. The second limitation of certain previous robustness studies is that they purposely attacked DNNs by finding specific perturbations that can alter model outputs. While these demonstrated the existence of robustness issue, the relevance to practical medical applications may be questionable, because the chance of having the specific perturbations may be small in real world. Instead, it is more meaningful to understand the robustness behavior under the naturally existing noise perturbations. 

With these considered, this study evaluates the robustness of a DNN-based classification model with combined simulation and experimental approaches in an example problem that classifies an axial CT image into one of four body sites. The practical relevance of this model is that such a model often serves as the basis for subsequent tasks. For instance, for organ segmentation task, the classification model can be used to determine the body site, before calling the corresponding segmentation model \cite{chen2021deep}. We trained this classification model using real patient data with whole body CT scans and the data from a realistic whole body CT phantom. This setup allowed us to repeatedly scan the CT phantom to understand the stability of the classification model with respect to naturally existing CT noises.   

\section{Methods}


\subsection{DNN-based Multi-class classification model construction}

\begin{figure}[htbp]
  \centering
  \includegraphics[width=0.9\textwidth]{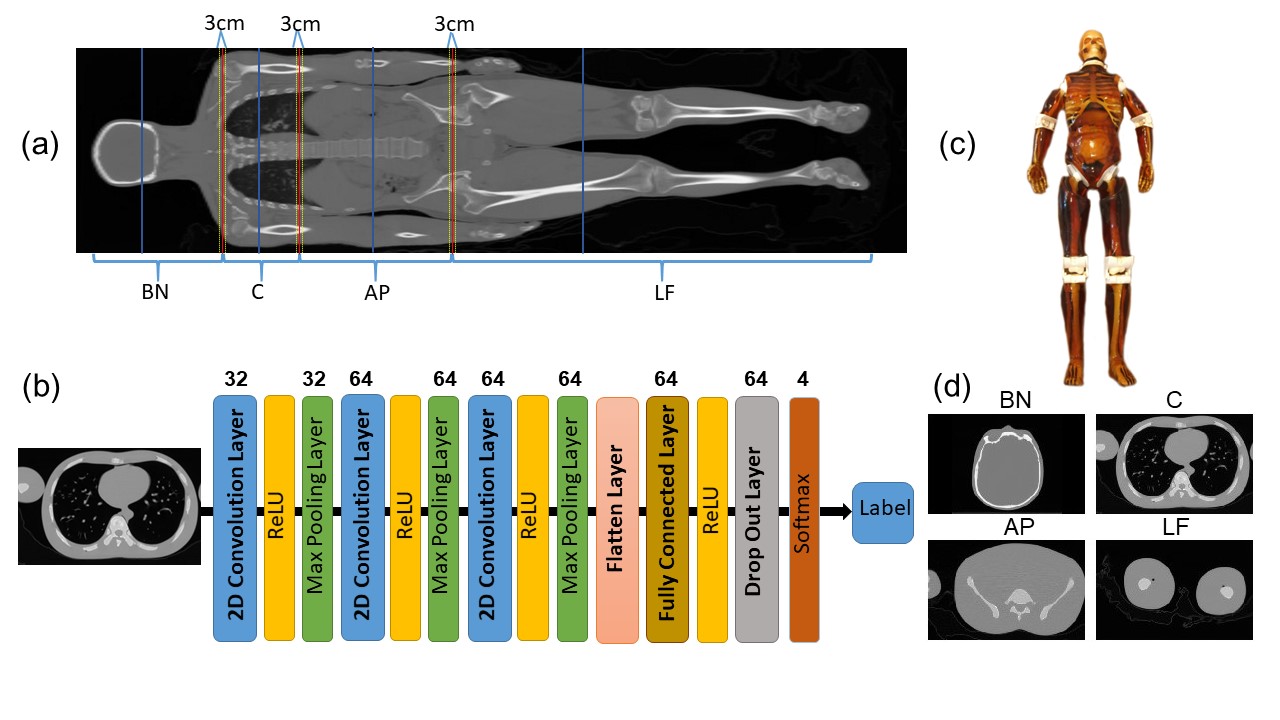}
  \caption{(a) Definition of the four body site regions, indicated by the red solid lines. Yellow dash lines are boundary regions removed when testing classification results. Blue solid lines indicate slices selected for experimental scans. (b) Structure of the DNN-based multi-class classification model. Numbers in each layer indicate the size of each layer in the constructed model. (c) Image of the orange-man phantom. (d) CT images of the phantom in experimental scans. }
 \label{Fig:model}
\end{figure}

We built a DNN-based multi-class classification model that classifies an axial CT image as one of four body sites: brain and neck (BN), chest (C), abdomen and pelvis (AP), leg and foot (LF). As indicated in Figure \ref{Fig:model}(a), from top to bottom, the BN site was defined as the region from the top vertex of the head to the upper lung limit. The C site included slides containing lung area. The AP site stopped at the top of femur head. The rest was defined as the LF site. We remark that the boundaries to separate adjacent regions were defined to some extent arbitrary for this study and by no means should be interpreted as accurate definitions in anatomy.

As shown in Figure \ref{Fig:model}(b), the DNN model followed a standard design for classification. The input was a 2D axial CT image. Three 2D convolutional layers were incorporated in the first half of the network. Each of the convolutional layers were followed by a rectified linear unit (ReLU) layer \citep{nair2010rectified} and a 2D max-pooling layer, to extract features from the input image. These features were then passed through one flatten layer and two fully connected layers. One of the fully connected layers was followed by a ReLU layer. The other was utilized with a softmax activation function for the classification purpose to decide the output label $y=1,\ldots,4$ for the four sites, respectively. Besides, we added a dropout layer before the last softmax layer to reduce overfitting \citep{sultan2019multi,srivastava2014dropout}. 

Training and testing were performed using 42790 axial CT images from whole-body CT scans of 47 patients each scanned once, and an orange-man phantom (Pure Imaging Phantoms, Farnham Royal, UK) (Figure \ref{Fig:model}(c) scanned for 20 times with different mAs levels. The whole body orange-man phantom is a human-size, full body anthropomorphic phantom with a state-of-the-art synthetic skeleton, lungs, liver, mediastinum and kidneys embedded in original soft tissue substitute. All the images were acquired on a Philips Brilliance CT scanner (Philips Healthcare, Amsterdam, Netherlands). $\sim$ 80$\%$ (34839 samples) of CT axial images were selected for training purpose, which included CT images of 37 patients and 10 CT scans on the orange-man phantom with various mAs levels. A ten-fold cross validation (CV) strategy was implemented. The 34839 axial CT images were split into ten groups of approximately the same size by patients/phantom. For each fold, one group was picked as validation set, while the remaining samples were used for model training. Data augmentation was performed to enlarge the training dataset by randomly scaling the CT images with factors in $80\sim 120\%$. 

The DNN models were trained by solving the following optimization problem:
\begin{equation}
\theta^*=\arg\min\limits_{\theta}\sum_{i\in T}\|D(x_i|\theta)-y_i^*\|_2^2, 
	\label{Eq:opt}
\end{equation}
where $x_i$ represents the $i$-th CT image in the training dataset. $y_i^*$ is the ground truth label for $x_i$. $D(x_{i}|\theta)$ represents the DNN function with $\theta$ being the network parameter set to be determined. The optimization problem was solved using the adaptive moment estimation (Adam) algorithm \citep{kingma2014adam} with Python with TensorFlow. The models were trained with 20 epochs using a batch size of 64 and an learning rate of $1\times10^{-4}$. These hyper-parameters were manually adjusted for the best performance. All the computations were implemented on a radiation oncology high performance computing server at UT Southwestern medical center.

After training, the model was tested using the remaining 20\% data (${N_{test}}$ = 7951) that were unseen by the training process. The performance was evaluated using metrics including precision, recall and F1-score for the specified classes. One issue in the evaluation was the to some extent ambiguously defined boundaries between body sites. To mitigate this impact, we also performed a test using a dateset by removing a transition zone at each boundary, as indicated by the yellow dashed lines in Figure \ref{Fig:model} (a)). The transition zone was defined as a region of six CT slices (3 cm) centering at each boundary slice. Among the models trained from the 10-fold CV study, the one with the highest F1-score averaging over the four classes in the independent test was advanced to the next step for robustness evaluations.

\subsection{DNN-based Multi-class classification model robustness evaluation}

Once the DNN model's performance was confirmed, we evaluated its robustness with respect to realistic CT noises in the input. We focused the study on the orange-man phantom, which allowed us to perform both numerical simulation and experimental studies. For the trained model, let $Y_{i}$ denote the model's predicted label for an unperturbed input axial CT image $X_{i}$. For each $X_i$, we could add realistic (via simulations or experiments) CT noises and fed the perturbed images into the DNN to output the new predicted label, denoted as $Y_{i}^{'}$. Evaluation of $Y_{i}^{'}$ for $X_i$ with noise added is called attack. An attack is considered successful, if $Y_{i} \neq Y'_{i}$. A reliable multi-class classification model is expected to be robust against noise perturbations. 

To quantify the model's robustness, a metric Successful Attack Rate (SAR) was introduced. For a given body site $i$, SAR was determined as:
\begin{equation}
	SAR_i=\frac{N_{succ}^i}{N_{attack}^i}(\%),
    \label{Eq:SAR}
\end{equation}
where $N_{succ}^i$ represents the number of successful attacks and $N_{attack}^i$ is the total number of attacks. The lower SAR is, the more robust the DNN model is. 

We also computed Confusion Matrix for Robustness (CMR) for the multi-classification model. Different from the confusion matrix commonly used to represent the percentage of an actual class being predicted to different classes, CMR represents the percentage of a predicted class without noise being predicted to different classes under noise. The matrix is of a size $4\times 4$ in this study corresponding to the four body sites. An identity matrix indicates the most robust case with the predictions unaffected by noise. Note that for each column $i$ corresponding to a given body site, sum of off-diagonal elements equals to $SAR_i$ by definition.


\subsubsection{Generating samples in simulation study}\label{simulation}
\

For numerical simulation, while our previous study used simulated CT noise to evaluate model robustness \citep{shen2020robustness}, the noise was generated for a single patch from a given noise power spectum (NPS). Yet it is known that CT noise is non-stationary and spatially correlated, the current study utilized an image domain noise insertion method to generate a whole image noise to investigate DNN robustness in a more realistic setting \citep{divel2019accurate}. In particular, we aimed at generating realistic CT noise for the orange-man phantom, so that we could compare numerical study results with those of experimental studies.

As such, we selected one slice of interest from each body region in the orange-man phantom denoted by the blue solid line in Figure \ref{Fig:model}(a). The slice of interest in each body site located in the middle region of each body site. This choice was made to allow us focusing on the evaluations of DNN model's robustness, as robustness of slices close to the boundary between two adjacent sites may be affected by ambiguity of defining a body site. For a CT slice of interest, we first repeatedly scan the phantom for 20 times at a reference mAs level of 25 mAs to estimate its NPS. The repeated CT images shared the same structure but different noise realizations. We obtained the average CT image and the noise realizations by subtracting the average image from each CT image. After that, we divided the image region into overlapping patches, within which the noise was assumed to be stationary. The size of patches was 6.8 mm $\times$ 6.8 mm, and they were overlapped by 30$\%$ to avoid discontinuities at the boundaries. The NPS in a patch was estimated using the ensemble average of the power spectrum from multiple acquisitions \citep{siewerdsen2002framework}, as
\begin{equation}
    S(f_x,f_y|x,y)=\frac{{\Delta_x}{\Delta_y}}{N_{x}N_{y}} \left<|\mathcal{F}(n(x,y)|^2\right>,
    \label{Eq:NPS}
\end{equation}
where $f_x$, $f_y$ are the frequencies in the Fourier space, $\Delta_x$ and $\Delta_y$ are pixel sizes, and $N_x$ and $N_y$ are the number of pixels. $n(x,y)$ is the noise signal of each acquisition, $\mathcal{F}$ denotes the Fourier transform, and $\left<.\right>$ indicates the ensemble average.

With the NPS calculated for each patch, we generated the spatially correlated CT noise across the whole CT image. Specifically, a white normal Gaussian noise was constructed for an image that has the same matrix size as the CT image. Then the white noise was filtered in the image domain by $s(x,y)=\mathcal{F}^{-1}{\sqrt{S(f_x,f_y)}}$, the inverse Fourier transform of the square root of the estimated NPS $S(f_x,f_y|x,y)$ in Eq.~(\ref{Eq:NPS}), to achieve the proper spatial correlation within each patch. Note that these small patches were overlapped with each other. To obtain the complete image of a CT noise, we merged noise data in overlapped regions between adjacent patches using a weighted scheme with a trapezoidal windowing function. Finally, we scaled the noise amplitude corresponding to a targeted mAs level with a factor $\alpha$ as following:
\begin{equation}
    \alpha = \frac{\sigma(mAs)}{\sigma(mAs_0)}=\sqrt{\frac{mAs_0}{mAs}},
     \label{Eq:mAs}
\end{equation}
to account for the difference in CT noise standard deviation $\sigma$ between the reference mAs level ($mAs_0$, 25 mAs in this study) and the targeted mAs level. 

Using this method, we generated 1000 noise images under each specific mAs level in a wide mAs ranging from 12.5 mAs to 500 mAs for each slice of interest. Adding the noises to the average CT image yielded simulated CT images with realistic noises.

\subsubsection{Generating samples in experimental study}
\

As it is not possible to repeatedly scan patients, we used the orange-man phantom as an alternative to evaluate the robustness of DNN-based models. As such, we repeatedly scanned the phantom at the four slices of interest. The mAs level affects noise amplitude and hence has a significant influence on the model prediction robustness. Generally speaking, increasing mAs increases radiation dose of the CT scan and decreases the amount of noise, hence improving the contrast resolution of the image. Balancing the dose with the contrast resolution required for interpretation must be considered when determining examination settings. In this study, we performed experimental scans under 200 mAs and 25 mAs levels. The 200 mAs level is a commonly used level in our clinic for high quality scans. The lowest mAs level available at our scanner is 25 mAs, which was hence chosen to study model robustness in a high-noise setting. The lowest mAs setting with the largest noise magnitude also represents a typical setup for the low dose CT scan, e.g. for screening purpose. For a particular mAs level, e.g. 25 mAs, the experimental CT scans were performed 150 times repeatedly on each of the four body sites, respectively. During the repeated scans, all the CT acquisition parameters and settings such as field of view, and slice thickness were kept the same.



\subsection{Robustness evaluations}
\

With the simulation and experimental data acquired, we fed the data to the DNN model. SAR and CMR were computed to evaluate model robustness. 

It is also necessary to compare the results from experimental scans and simulated CT noise images. This was achieved via a statistical test. Specially, each entry of the CMR represents the probability $p_{i,j}$ of a given class $j$ predicted for noiseless CT images being predicted to a class $i$ under noise perturbations, for $i,j=1,\ldots, 4$. They was estimated based on repeated experiments via numerical simulations as $\hat{p}^1_{i,j}$ or actual CT scans as $\hat{p}^2_{i,j}$, but the corresponding true values $p^1_{i,j}$ and $p^2_{i,j}$ were unknown. We performed the test with the null hypothesis that $p^1_{i,j}=p^2_{i,j}$. Assuming the DNN prediction results under noise perturbations follow a binomial distribution, and under normal approximation to binomial distribution given the relatively large sample sizes, we computed $Z$ value as $Z_{i,j}= \frac{p^1_{i,j}-p^2_{i,j}}{\sqrt{p_{i,j}(1-p_{i,j})(\frac{1}{n_1}+\frac{1}{n_2})}} $, where \( p_{i,j}= \frac{n_1p^1_{i,j}+n_2p^2_{i,j}}{n_1+n_2} \), and $n_1=1000$ and $n_2=150$ are the number of experiments in simulation and phantom studies, respectively. We may reject the null hypothesis, if $|Z_{i,j}|$ is larger than the critical region value $Z^*= 1.96$ with 95\% confidence level in this two tailed test problem.

In addition, we also studied the robustness issue in two other aspects. 

\subsubsection{Robustness of models trained under the same training scheme}\label{section:sametraning}
\

The training process for a DNN model often inherently contains randomness. Even with the same training data and same hyper-parameter setting, the randomness occurs due to stochastic operations during the training process, such as the stochastic gradient descent method that randomly select a subset of training data at each optimization step. As such, the resulting models are not the same, and it is hence important to understand the robustness distribution of these models. For this purpose, we performed an evaluation study on the robustness of multiple models trained under the same setting. We repeatedly trained the multi-class classification model for 100 times using exactly the same training procedure, settings, as well as the training and validation dataset. We then evaluated the robustness of these 100 models using experimentally acquired CT images.

\subsubsection{Adaptive training approach to improve robustness}
\

In a previous study, we proposed an adaptive training scheme that improved the DNN-based models' robustness \citep{shen2020robustness} by repeatedly fine-tuning the DNN-based model. At each iteration, the scheme evaluated robustness of the trained DNN using simulated data with realistic noise perturbations, and then retrained the DNN model by adding to the training dataset those perturbations that successfully altered the DNN’s output. While our previous study demonstrated the effectiveness of this approach, only simulation studies were conducted. Here, we studied the effectiveness of the adaptive training scheme using experimental data. Specifically, we followed the proposed adaptive training scheme using simulation data to generate perturbations and retrain the model. In parallel to this process, we used the experimental data to measure robustness of the model and monitor the improvement. This is expected to serve as a ground truth way to show the validity of the adaptive training scheme.

\section{Results}

\subsection{DNN model's performance}\label{DNNmodelPerformance}
\

While model performance was not the focus of this study, we presented it here to ensure that subsequent robust analysis was built on models with reasonable performance. The precision, recall and F1-score for the DNN-based multi-class classification models obtained in the 10-fold CV scheme are listed in Table \ref{Table:DNNtrain2}. The results are the average and standard deviations over the ten models, showing their excellent performance. Testing 1 and 2 are independent tests using the whole testing data, and using the data without the transition zone between neighboring body sites, respectively. Note that the performance in Testing 2 was better than that of Testing 1, because of the removal of ambiguity of anatomy at the transition zones. Among the 10 models in the CV scheme, the model with the highest F1-score averaging over the four classes for the Testing 2 dataset was used to evaluate the robustness in the remaining of this paper.

\

\begin{table}[bhpt]
\centering
\caption{Multi-class classification performance on training, validation and testing datasets. Each result is average and standard deviations estimated over the ten-fold CV study. All numbers are in $\%$. }
\
\begin{tabular}{c C{1.9cm}C{1.9cm}C{1.9cm}C{1.9cm}C{1.9cm}}
\hline\hline
Datasets & Metrics & BN & C & AP & LF \\
\hline
\multirow{3}{*}{Training} & Precision  & 99.8$\pm$0.3 &99.5$\pm$0.4  &99.8$\pm$0.2 &99.9$\pm$0.0 \\
& Recall \hspace*{1.5mm}& 99.8$\pm$0.2 &99.6$\pm$0.4 &99.9$\pm$0.1 &100$\pm$0.0  \\
& F1-score  & 99.8$\pm$0.1 &99.5$\pm$0.2 &99.9$\pm$0.0 &100$\pm$0.0 \\
\hline
\multirow{3}{*}{Validation} &Precision &97.9$\pm$1.9 &95.9$\pm$2.4 &98.1$\pm$1.2 &99.4$\pm$0.7 \\
& Recall \hspace*{1.5mm} &98.5$\pm$1.3 &96.8$\pm$1.7 &98.1$\pm$1.5 &99.2$\pm$0.9 \\
& F1-score  &98.2$\pm$1.1 &96.3$\pm$1.2 &98.1$\pm$0.9 &99.3$\pm$0.4 \\
\hline
\multirow{3}{*}{Testing 1} & Precision  &99.3$\pm$0.7&96.0$\pm$1.2&98.1$\pm$0.6 &99.7$\pm$0.2 \\
& Recall \hspace*{1.5mm} &98.5$\pm$0.6 &95.8$\pm$1.7&98.8$\pm$0.4 &99.5$\pm$0.2  \\
& F1-score &98.9$\pm$0.2 &95.9$\pm$0.5&98.5$\pm$0.2&99.6$\pm$0.1\\
\hline
\multirow{3}{*}{Testing 2} & Precision &99.7$\pm$0.5 &99.7$\pm$0.3 &99.5$\pm$0.4 &100$\pm$0.0\\
& Recall \hspace*{1.5mm} &100$\pm$0.0 &98.0$\pm$1.1 &100$\pm$0.0 &100$\pm$0.0 \\
& F1-score  &99.8$\pm$0.3 &98.8$\pm$0.5 &99.7$\pm$0.2 &100$\pm$0.0\\

\hline\hline
\end{tabular}
\label{Table:DNNtrain2}
\end{table}

\subsection{Noise generation}

\begin{figure}[bt]
   \centering
  \includegraphics[width=0.9\textwidth]{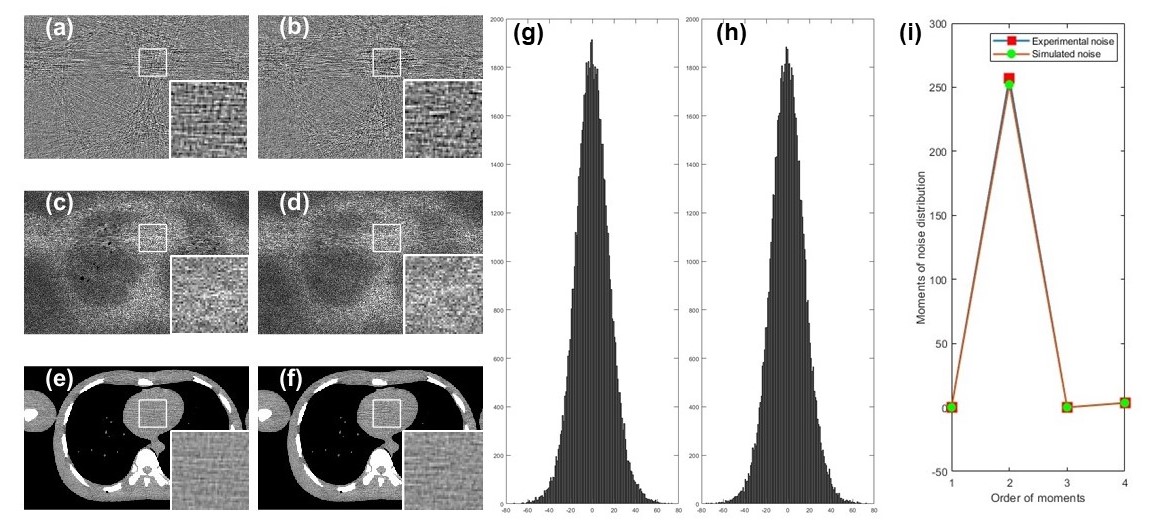}
  \caption{(a)-(b) CT noise images for CT scans and simulated noise under 25 mAs. Display window is [-80, 80] HU. (c)-(d) CT noise standard deviations for CT scans and simulated noise under 25 mAs. Display window is [0 40] HU. The white boxes are zoomed-in regions. (e)-(f) Experiment CT image and image with simulated noise under 25 mAs. Display window is [-200 200] HU. (g)-(h) Histograms of noise images for CT scans and simulated CT images. (i) Comparisons of moments of noise distributions for CT scans and simulated CT images. }
 \label{Fig:chest_noise}
\end{figure}

We first demonstrated validity of the simulation approach to generate realistic noises by comparing the noise images and noise standard deviation images obtained in actual CT scans and simulations. As shown in Figure \ref{Fig:chest_noise} (a-f), the simulated noise possessed visually a similar pattern as the noise in the experimental CT scan. The standard deviation images were calculated over 20 experimental CT noises and 20 simulated realistic noises. They both exhibited close correlations with the underlying anatomical features, as well as similar variance maps. It is difficult to quantitatively measure the statistical agreement between the simulated and actual noise signals. As a means for demonstration, we plot histograms of the noises in Figure \ref{Fig:chest_noise} (g) and (h). We also computed up to the 4th order of moment of the two distributions as in Figure \ref{Fig:chest_noise} (i). The differences were within $2\%$, reflecting agreement between the two. 

\subsection{Robustness evaluations}

\subsubsection{Robustness of trained model}

\begin{figure}[htbt]
   \centering
  \includegraphics[width=\textwidth]{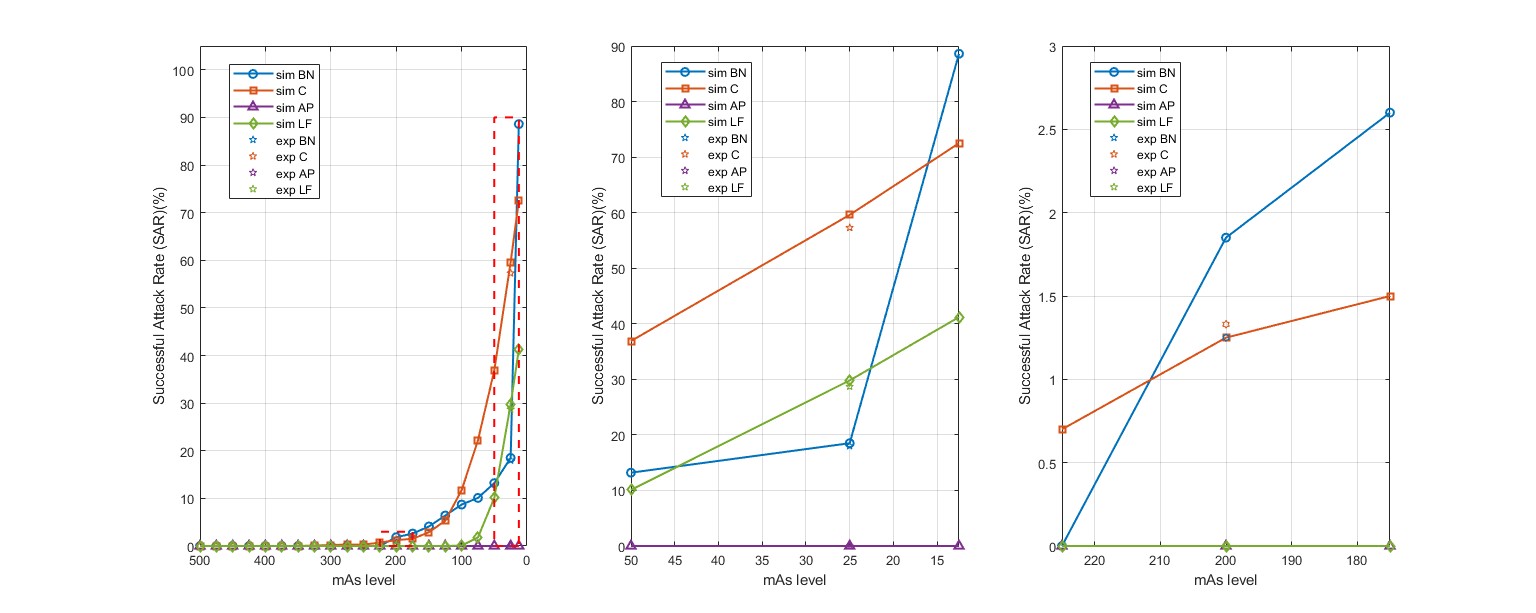}
  \caption{(a) SAR with respect to CT noises under different mAs levels (from 12.5 mAs to 500 mAs). Marker stars denoted the SAR for experimental studies under 200 mAs and 25 mAs, respectively. Rectangular boxs indicated the zoomed-in regions in (b-c). (b) Zoom-in view of plots for mAs range in [12.5, 50] mAs. (c) Zoom-in view  of plots for mAs range in [175, 225] mAs.}
 \label{Fig:SAR}
\end{figure}

In the simulation study, we generated 1000 simulated noise images at each of the four body sites for the orange-man phantom under different mAs levels ranging from 12.5 mAs to 500 mAs. SAR of the DNN-based model for the four body sites were calculated and the results are summarized in Figure \ref{Fig:SAR}(a). First, generally speaking, the robustness of DNN output results became worse with reduced mAs level. This is expected, as lowering the mAs level resulted in increased noise levels and hence large perturbations to the input CT images that tend to increase the chance of changing the DNN's outputs. Second, different classes attained robustness to different degrees. The class AP was found to be the most robust with respect to noise perturbations. All the predicted results kept unchanged with noise perturbations, and SAR for this class was zero. However, the other three classes (BN, C and LF) showed significantly weaker robustness. In Figure \ref{Fig:SAR} (b) and (c), the plots of SAR were displayed in zoomed-in views under the mAs ranges from 50 mAs to 12.5 mAs, and 225 mAs to 175 mAs, respectively. It was observed that under 200 mAs, the LF class was robust against the noise perturbations, with SAR being zeros. As mAs was reduced, it became sensitive to noise perturbations starting from 100 mAs level, where SAR started to be non-zero. For BN and C classes, robustness issue started to appear from 225 mAs and 325 mAs, respectively. At the large noise limit of 12.5 mAs, SAR of BN, C, and LF were 88.6\%, 72.5\%, and 41.2\%, respectively.

As for the experimental studies, the results shown as star makers in Figure \ref{Fig:SAR} generally agreed with the simulation results. At 200 mAs level, SAR for the BN and C sites both were 1.25\%, while for the AP and LF sites, SAR were zeros. However, under 25 mAs level, SAR for the BN, LF, and C sites increased to 18\%, 28.7\%, and 57.3\%, while that for the AP site remained at zero. These experimentally obtained SARs were consistent with those computed by the simulated CT noise perturbations, as shown in Figure \ref{Fig:SAR}. 

\begin{figure}[h]  
\centering 
\includegraphics[width=1\textwidth]{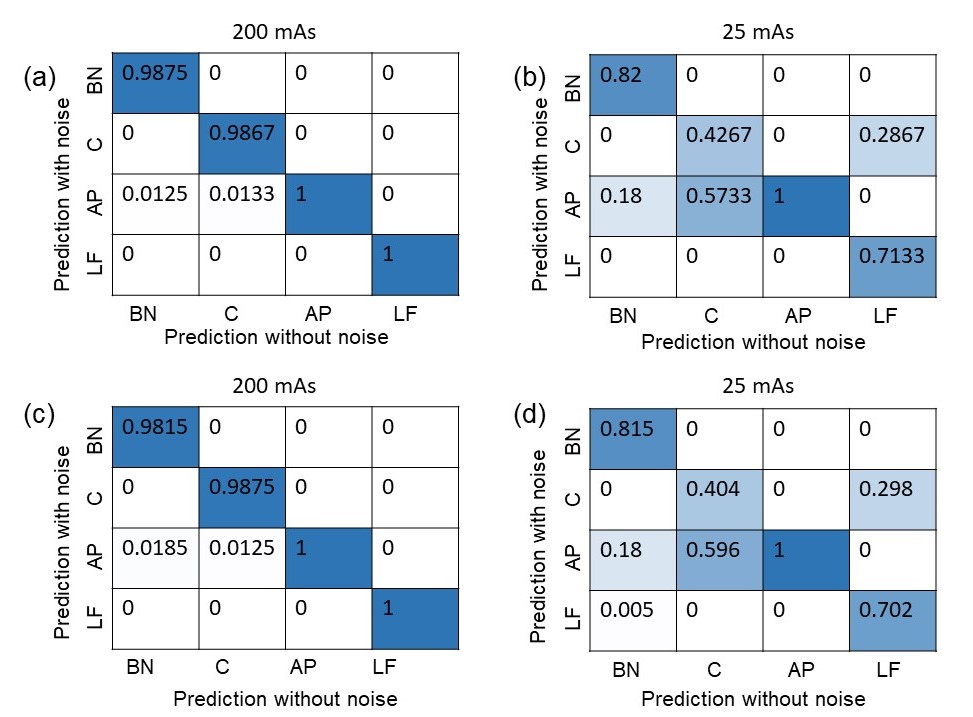}
\caption{CMR of (a) experimental study at 200 mAs; (b) experimental study at 25 mAs; (c) simulated study at 200 mAs; and (d) simulated study at 25 mAs.}
\label{Fig:cm}
\end{figure}

In Figure \ref{Fig:cm}, the CMRs are presented for experimental and simulation studies at 200 mAs and 25 mAs. At 200 mAs, the C and BN sites appeared to be slightly not robust. For instance, in the experimental study, majority of the C site was still predicted as the C site with noise perturbations, but with 1.33\% probability the prediction was incorrectly as AP. This number was close to the simulation result in Figure \ref{Fig:SAR}, where SAR of simulated CT chest axial images at 200 mAs was 1.25\%; while under 25 mAs, 86 of 150 CT chest scans were predicted as AP site, resulting in the experimental SAR value of 57.33\%, which was close to the simulated SAR value of 59.6\%. 
 
\begin{figure}[bt]
   \centering
  \includegraphics[width=0.7\textwidth]{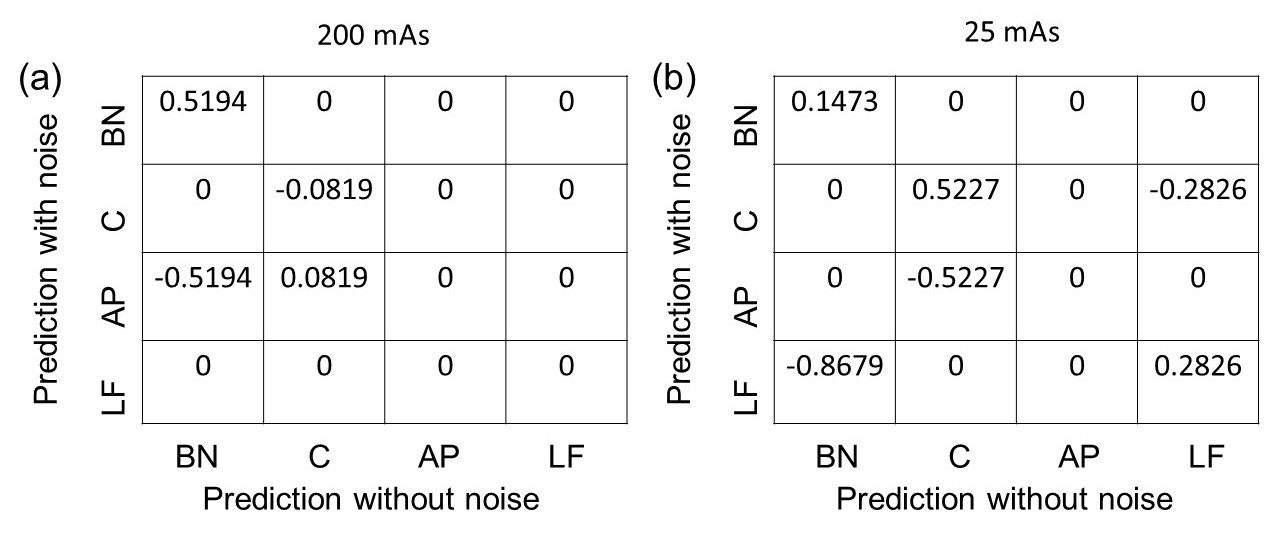}
  \caption{Test statistics of the null hypothesis testing that the calculated CMRs of experimental scans and simulated CT noisy images were statistically same with each other under (a) 200 mAs and (b) 25 mAs.}
 \label{Fig:cm_hypothesis}
\end{figure}

To compare the experimental and simulation results, Figure \ref{Fig:cm_hypothesis} presents $Z$ values for 200 mAs and 25 mAs. All the $|Z|$ values were less than the critical value $Z^*=1.96$. Hence, at the 95\% confidence level, we cannot reject the null hypothesis testing that the corresponding entries in CMRs calculated by experimental studies and by numerical simulations are statistically the same. 



\subsection{Robustness of models trained under the same training scheme}
\ 

As described in Section \ref{section:sametraning}, we repeatedly trained the DNN-based classification models for 100 times using the same setup. The precision, recall and F1-score of each DNN model obtained for the testing dataset without axial CT images at boundaries between adjacent body site regions (Testing2 in Table \ref{Table:DNNtrain2}) were calculated. The mean value of precision over the 100 trained DNN models for BN, C, AP and LF classes were 99.6\%, 99.4\%, 99.1\% and 98.8\%, the averaged recall were 100\%, 98.6\%, 99.9\% and 99.5\%, and the averaged F1-score were 99.9\%, 98.8\%, 99,6\% and 99.2\%, respectively. These high values showed excellent performance of the 100 DNN-based models on multi-class classification for the testing dataset. 

\begin{figure}[bt]
   \centering
  \includegraphics[width=0.8\textwidth]{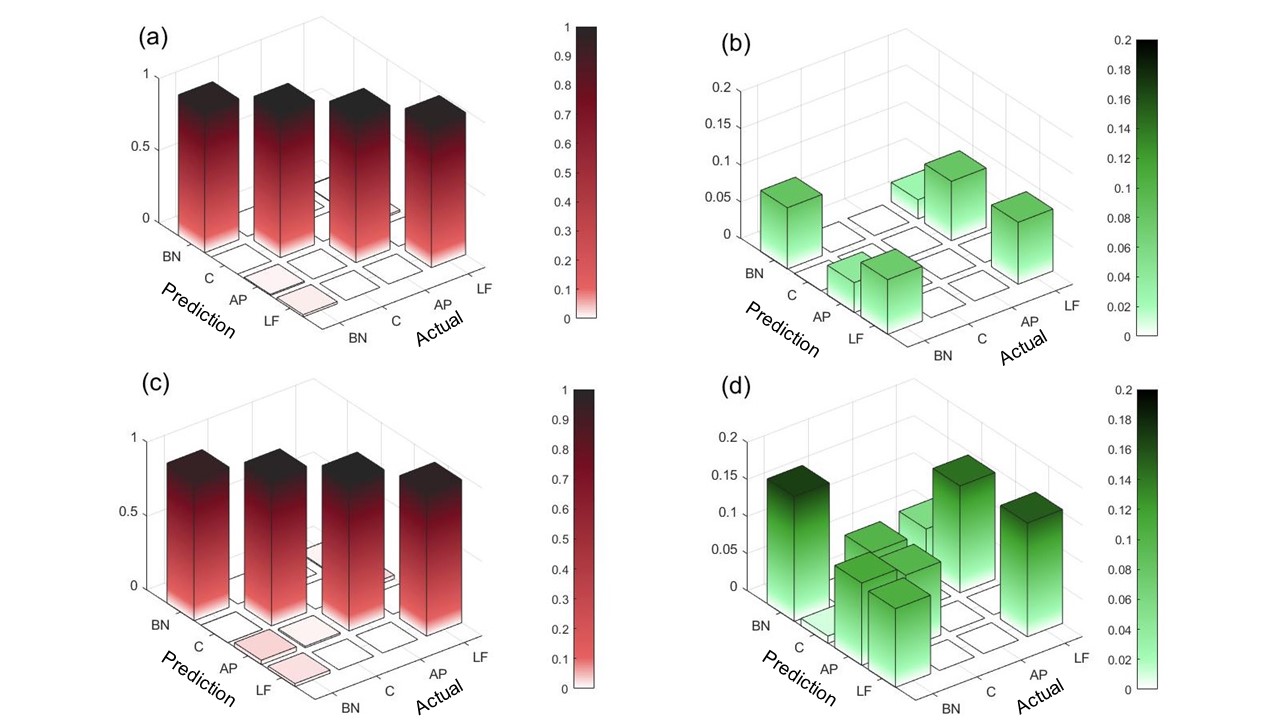}
  \caption{3D view for (a) mean and (b) standard deviation of CMRs of 100 DNN-based models against CT scans under 200 mAs; (c) mean and (d) standard deviation of CMRs of 100 DNN-based models against CT scans under 25 mAs.}
 \label{Fig:100models}
\end{figure}

Subsequently, we performed the robustness study on the 100 well-trained DNN models using the experimental scans. The results are shown in Figure \ref{Fig:100models}. Out of the 100 DNN models, 14 models exhibited robustness issue with non-zero SAR for BN, C and LF sites under 25 mAs. We summarized the CMRs of all the 100 models by presenting the element-wise mean values and standard deviations. As seen in Figure \ref{Fig:100models} (a) and (c), both of the averaged CMRs under 200 mAs and 25 mAs showed non-zero off-diagonal elements, with the amplitudes of off-diagonal elements increased with noise. SAR for the BN, C, AP and LF sites at 200 mAs were 2.37\%, 0\%, 0, and 1.59\%, respectively; and those at 25 mAs were 5.21\%, 1.23\%, 0\%, and 3.36\%, respectively. Figure \ref{Fig:100models} (b) and (d) showed the standard deviations of 100 CMRs. Under 25 mAs, the standard deviations of CMRs were larger than those under 200 mAs. 
These results demonstrated that even though the 100 DNN-based models were trained in exactly the same training scheme and dataset, because of the inherent randomness during the training process, the results had different level of robustness. Similar to the previous findings, these 100 models on exhibited different robustness extent on the four classes. Among the four body sites, prediction on the AP site was found to be much more robust than the other three sites.

\subsection{Adaptive training scheme to improve robustness}
\

\begin{figure}[bt]
    \centering
    \includegraphics[width=1\textwidth]{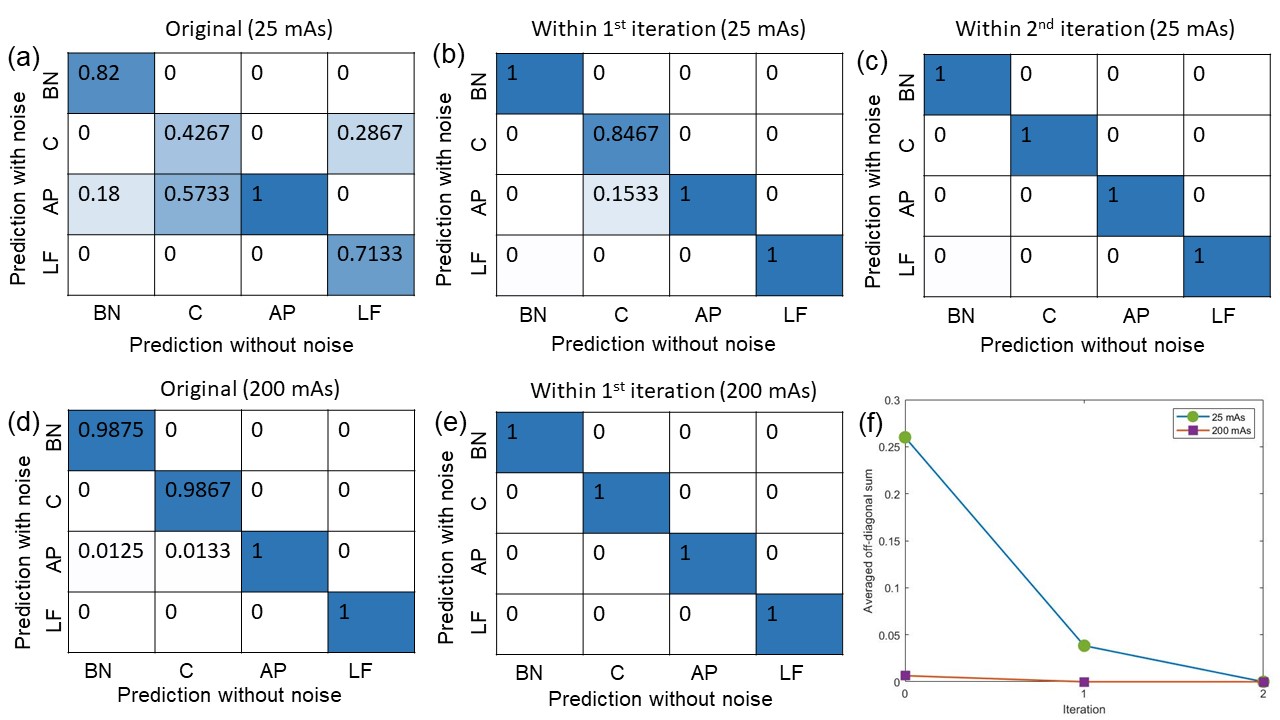}
\caption{(a)-(c) CMRs during the adaptive training process for 25 mAs level. (d)-(e) CMRs during the adaptive training process for 200 mAs level. (f) Averaged off-diagonal sums over all sites during the adaptive training process.}
 \label{Fig:cm_adaptive}
\end{figure}

We employed the adaptive training scheme to improve the DNN models’ robustness in this study. Experimental study results on the robustness of the adaptively trained DNN models are presented in Figure \ref{Fig:cm_adaptive}. As the perturbed samples that successfully attacked the trained network were added to the training dataset, the newly trained DNN model showed significantly improved robustness. Specifically, for the 200 mAs case, SAR for BN and C sites were reduced to zeros after the adaptive training scheme was applied with one iteration. For the 25 mAs case, SAR for BN and LF sites were reduced to zeros at the first iteration, while SAR for C site was reduced to 15.3\% from 57.3\%. After two iterations, SARs for all sites became zero. In Figure \ref{Fig:cm_adaptive} (f), the averaged SARs over all sites are plotted against the number of iterations for the two mAs levels. The monotonically decaying trend of the two curves indicated validity of the adaptive training scheme to improve model robustness. 

Additionally, we further tested the adaptively trained DNN models after two iterations using the numerically simulated CT images at four body sites under mAs levels from 500 mAs to 12.5 mAs. It was found that SARs for all the considered cases were reduced to zeros.

\section{Discussions}
\

Robustness is a critically important aspect to consider, when developing DNN models for clinical applications. Our group, and others, have previously investigated robustness of DNN models in extensive studies \citep{shen2020robustness,gao2021improving,cui2021deattack,antun2020instabilities,wu2020stabilizing,wu2021drone}. Compared to existing studies, the current work is different in the following aspects, which hence highlights the contributions of this study. 

First, this study employed the simulation approach to study the robustness of DNN-based multi-class classification models by generating the simulated realistic CT noises in the entire image domain. While our previous study \citep{shen2020robustness} generated CT noises via simulation for robustness evaluation, noise was conducted at a patch level, and hence the simulated CT noise ignored position dependent of NPS, a well-known property of the CT image noises. The current study overcame this limitation by generating noise images considering NPS variations through the image domain. Our experimental studies validated the accuracy of generated noise signal.

Second, to our knowledge, this is the first time that a comprehensive study including both simulation and experimental investigations were performed to demonstrate the robustness issue of a DNN-based classification model in a clinically relevant problem. We conducted the study with repeatedly scans of the orange-man phantom at a few slices under various mAs levels. The robustness measure in terms of SAR and CMR were found to be quantitatively in agreement between the simulation and the experimental studies. The experimental study served as a ground truth approach to demonstrate robustness issue of a well-trained DNN classification model, as well as to show the validity of simulations to investigate the robustness issue.

Third, this study focused on naturally existing noise in CT images. Robustness concern DNN-based classification model has been long studied in general. Yet, although it has been demonstrated in different contexts, most existing studies focused on attacking DNN models, namely to purposely identify perturbations that can manipulate DNN model outputs \citep{eykholt2018robust,laugros2019adversarial,akhtar2018threat,yuan2019adversarial,finlayson2018adversarial}. While those studies undoubtedly showed existence of perturbations capable of affecting DNN model performance, the clinical relevance may be questioned, as the purposely constructed perturbations may not exist at a high chance. Different from these studies, the perturbations in this work were sampled from the actual CT image noise domain, via simulations or experiments. The robustness concern of the trained DNN model under these noise perturbations made our study more relevant to real clinical applications. 

Last, in addition to evaluating robustness for one trained model, we also investigated variations of robustness among models trained under the same setting. The variation of robustness is a valid concern. Due to randomness in model training caused by the stochastic nature of the numerical algorithm, while the resulting trained model all achieved a satisfactory level of classification performance, we found that they may behave with different levels of robustness. Ultimately, a DNN model represents a highly non-linear function mapping between an input CT image and a classification label, whose output may be sensitive to input image values. Due to the highly non-convex optimization problem of DNN model training, the resulting DNN models under repeated training runs are local minima of the objective function, which attain different numerical properties. Our study revealed the variation of robustness among them, hence highlighting the need for assessing model robustness on a case-by-case bases.

Meanwhile, our study has the following limitations. As DNNs have been widely used in a broad scope with different network setups for applications etc, we would like to remark on the validity of the study to avoid over-interpreting the results. 

First of all, while using the phantom allowed us to demonstrate the robustness issue via a real experimental setting, strictly speaking, our study only showed the existence of the robustness concern, when classifying a CT slice of the phantom. The existence of this problem in real patient case was still unverified. Yet we hope the realistic appearance of the orange-man phantom relative to real humans in CT image can support the use of the phantom as a surrogate to investigate the robustness problem for real patient cases. Moreover, as our study, e.g. the statistical test, showed validity of using simulations as a means for robustness study, we will perform more studies with simulated noise perturbations to comprehensively investigate the problem in patient cases. 

Another limitation is the relevance of the selected classification problem, i.e. body site identification, to the actual clinical problems of interest, for instance lung nodule classification. As mentioned previously, we chose the body site identification problem for the purpose to allow experimental studies on a physical phantom. Yet, for this classification task, it is expected that the trained DNN model may extract image features at the body scale, such as cross section size and shape, to decide the classification label. The feature selection may not be of relevance to lung nodule classification problem, which focuses on small scale features for the nodules. Hence, the robustness concern identified from this study may not be generalizable to the lung nodule classification problem. We expect that phantom studies with realistically 3D printed nodules may be performed to evaluate robustness issue for the lung nodule classification problem \citep{hatamikia2023realistic}, although it may be a challenge to replicate nodules of different categories.

Third, this work only studied the robustness on one network model with a very specific and relatively simple network structure for multi-class classification purpose. It is expected that the robustness level depends on many factors such as specific model structure and model training process. Hence, it should be understood that the observed behaviors are for the model studied here, whereas generalization of the discoveries to other models may not hold. With the rapid progress of DL, advanced models have been explored, such as transformer \citep{willemink2022toward}, diffusion model \citep{kazerouni2022diffusion} etc. There have been initial studies on the robustness of these models. It is our ongoing work to evaluate these models' robustness using the simulation and experimental approaches developed in the current study.

\section{Conclusion}


In this paper, we investigated robustness of a multi-class classification DNN-based model with respect to CT noises via simulation and experimental approaches. The results in experimental study agreed with those in simulations, both showed that the DNN model had robustness issue to CT noises to the degree depending on mAs levels. We also demonstrated that repeatedly trained DNN models with exactly the same training scheme attained different levels of robustness, and that the adaptive training scheme is effective in terms improving robustness of the DNN model.

\section*{Acknowledgement}
This work was funded in part by the National Institutes of Health under the grants R37CA214639, R01CA227289, R01CA237269, R01EB032716, and R01CA254377.

\bibliography{reference}

\end{document}